# Direct x-ray scattering signal measurements in edge-illumination/beam-tracking imaging and their interplay with the variance of the refraction signals

Ian Buchanan[1]*, Silvia Cipiccia[1], Carlo Peiffer[1], Carlos Navarrete-León[1], Alberto Astolfo[1], Tom Partridge[1], Michela Esposito[1], Luca Fardin[2†], Alberto Bravin[2‡], Charlotte K Hagen[1], Marco Endrizzi[1], Peter RT Munro[1], David Bate[3], Alessandro Olivo[1]

[1]Department of Medical Physics and Biomedical Engineering, University College London, Gower Street, WC1E 6BT, UK

[2]European Synchrotron Radiation Facility, Grenoble, 38043, France

[3]Nikon X-Tek System Ltd., Tring Business Centre, Tring, Herefordshire, UK

*Corresponding author: ian.buchanan.15@ucl.ac.uk

†current address: University of Grenoble Alpes, Strobe Inserm UA7, Grenoble, France

‡current address: University Milano Bicocca, Dept. of Physics "G. Occhialini", Milano, Italy

## Abstract

X-ray dark-field or ultra-small angle scatter imaging has become increasingly important since the introduction of phase-based x-ray imaging and is having transformative impact in fields such as *in vivo* lung imaging and explosives detection. Here we show that dark-field images acquired with the edge-illumination method (either in its traditional double mask or simplified single mask implementation) provide a direct measurement of the scattering function, which is unaffected by system-specific parameters such as the autocorrelation length. We show that this is a consequence both of the specific measurement setup and of the mathematical approach followed to retrieve the dark-field images. We show agreement with theoretical models for datasets acquired both with synchrotron and laboratory x-ray sources. We also introduce a new contrast mechanism, the variance of refraction, which is extracted from the same dataset and provides a direct link with the size of the scattering centres. We show that this can also be described by the same theoretical models. We study the behaviour of both signals vs. key parameters such as x-ray energy and scatterer radius. We find this allows quantitative, direct, multi-scale scattering measurements during imaging, with implications in all fields where dark-field imaging is used.

## Main

X-Ray dark field (DF) imaging has attracted significant interest over recent years, thanks to its ability to detect the presence of features below the spatial resolution of an imaging system[1–5]. It belongs to the class of "phase-sensitive" x-ray imaging methods[6], in which image contrast arises from the unit decrement of the real part ($\delta$), rather than from the imaginary part ($\beta$) of the refractive index. "Ultra-small angle x-ray scatter" (USAXS) is often used as a synonym, with the caveat that scattering angles involved in DF are in the region of microradians or sub-microradian, while USAXS usually refers to angles of a fraction of a degree. Following its introduction at synchrotron facilities in crystal-based experiments[1–3], DF became increasingly popular after Pfeiffer *et al*'s demonstration that the same signal could be accessed with conventional laboratory sources by using a three-grating setup[4]. Thereafter, DF with laboratory sources was demonstrated with Edge-Illumination[5] (EI) and its simplified, single mask "beam tracking" (BT) implementation[7], as well as other methods such as speckle-based imaging[8,9]. Following initial applications to e.g. welding[10] and composite materials[11], DF has significantly evolved and is currently being used for *in vivo* lung imaging of human patients[12]. A recent study also showed significant potential in the detection of concealed explosives[13], and efforts to implement DF on a clinical CT gantry are underway[14]. Among the many phase-based techniques, we focus here on EI[15] (and on its BT implementation, which earlier work demonstrated to provide equivalent signals[16]), as it is achromatic[17] and can be implemented with low spatial coherence[18].

A straightforward way to understand the nature of the DF signal is by referring to x-ray refraction. This is greatest at sample boundaries (as the refraction angle is proportional to the first derivative of $\delta$ in space[6]) and after some propagation, directional changes translate into a change in the (transverse) position at which the x-rays are detected. These changes are usually referred to as "phase contrast" when they are resolved by the imaging system, and as DF when they are not. Refraction, modelled through ray-tracing, provides an approximation of phase-based mechanisms; wave-optics tools (e.g. Fresnel-Kirchhoff diffraction integrals) are required for a more accurate description[6]. Differences are significant only when the object is coherently illuminated; under limited coherence conditions (e.g. those of a laboratory x-ray source), the refraction (i.e. ray-tracing) and wave-optics descriptions provide compatible results[19]. The former is simpler and lends itself to straightforward implementation in Monte-Carlo approaches.

We implemented both ray-tracing Monte-Carlo and wave-optics models without resorting to the thin object approximation. Concurrently, we demonstrate that the same results are obtained using classical scattering theory when multiple scattering is negligible. Details on the two models are provided in the methods section.

The intensity of scattered radiation can be measured as a function of the scattering momentum vector, $\boldsymbol{q}$ (where $|\boldsymbol{q}| = \frac{4\pi}{\lambda} \sin(\theta)$, $\lambda$ is the x-ray wavelength, and $2\theta$ is the scattering angle). For a thin layer of uniformly distributed spheres, the intensity takes the form[20]:

$$I(q) = |\Delta\rho|^2 V_f \left| \frac{3(\sin(qR) - qR \cos(qR))}{(qR)^3} \right|^2,$$

(1)

Where $\Delta\rho$ is the difference in electron density between the spheres and the surrounding medium, $V_f$ is the volume of spheres, and $R$ is the sphere radius; we also note that $I(0) = |\Delta\rho|^2 V_f$.

Scattering measurements can refer to wide angles on the order of several degrees (molecular; crystalline), small angles on the order of arc seconds (complex biological materials; soft matter), and to ultra-small angles on the scale of microradians (microscopic or sub-microscopic structures)[21]. In wide and small angle scattering experiments, the primary beam is often stopped, as the characteristic peaks are of such low intensity that, if the primary beam were integrated alongside them, detectors would saturate quickly. In USAXS as measured in EI/BT and other phase methods, the higher order peaks of $I(q)$ are not measured, and the broadening of the primary beam becomes the object of analysis.

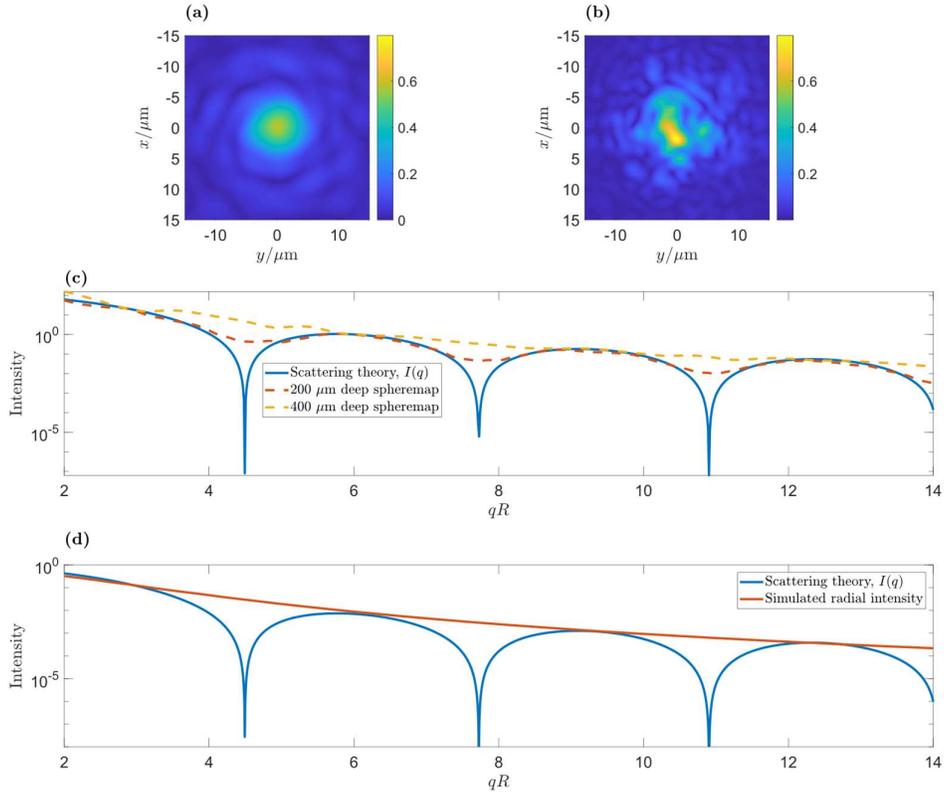

*Fig 1 Wave-optics and Monte-Carlo modelling of EI/BT vs scattering theory. Panels (a) and (b) show wave-optics results for a pencil beam scattered by a thin (200 μm) and a thick (400 μm) layer of 1 μm diameter microspheres, respectively, with the colour bars showing X-ray intensities relative to the peak of an unperturbed probe beam. Radially integrated plots are overlapped to the profile predicted by scattering theory (solid blue line) in panel (c): as can be seen, intensity from the thin layer (dashed red line) matches it much better than the that from the thick one (dashed yellow). Panel (d) shows that ray-tracing cannot reproduce the intensity oscillations and only provides the curve's envelope; however, in conditions of limited coherence and/or multiple scattering, the three models are shown to be compatible.*

Nonetheless, by using the wave-optics model developed for EI/BT (see methods), we can demonstrate both adherence to and departure from Eqn. (1) under conditions of single and multiple scattering, respectively. Fig. 1(a) and (b) display the intensity field resulting from a 15 keV plane wave interacting with sphere ensembles of constant packing density in 200 μm and 400 μm thick slabs. Fig 1(c) shows that the (radially-integrated) intensity resulting from interactions with the 200 μm thick sample approximate Eqn. (1) well (aside from the difficult-to-sample deep troughs), while the intensity oscillations tend to disappear in the case of multiple scattering caused by the thicker sphere ensemble. From the model's experimental validation (see below), these results demonstrate the link between scattering theory and EI/BT-based USAXS measurements. Fig 1(d) shows the radial intensity from the same sample as panel (b) modelled with Monte-Carlo ray tracing, which only models photon intensities rather than amplitudes, and thus does not reproduce the oscillations caused by coherence effects but, rather, traces the envelope of the function. However, as shown by the dashed yellow line in Fig. 1(c), these same oscillations are washed out by multiple scattering in thick samples like those analysed in this work. This demonstrates the link, and underlying consistence under the conditions of limited coherence and/or multiple scattering between classic scattering theory, wave-optics and Monte-Carlo modelling.

In EI/BT, one or more fan beams (hereafter, 'beamlets') are used instead of pencil beams, making them full-field imaging methods (or 1D scanned methods when a single beamlet is used). The observed change in a beamlet's intensity, centre of mass and width correspond to transmission, refraction, and DF measurements, respectively[22]. Sample properties are inferred via the combination of these, quantitatively supported by forward modelling which was shown to agree with experimental data[23,24]. DF contrast is the width of the sample-induced distribution of refraction angles[25]: the greater the average number (and magnitude) of refraction events within a given beamlet, the broader the emerging beamlet i.e., the greater the DF signal. This is expressed as the change in angular variance of beamlet intensity, $I(x)$, with the sample present, minus the reference variance without the sample:

$$M_2^{S,R} = \frac{1}{M_0^{S,R}} \int (x - M_1^{S,R})^2 I^{S,R}(x) \, dx \quad (2a)$$

$$DF = (M_2^S - M_2^R)/z_2^2 \quad (2b)$$

In Eqn. (2a), we are utilising the moments approach to EI/BT data processing[25], with superscripts $S$ and $R$ referring to sample and reference beamlets, respectively. DF signals are the change in second moment, $M_2$, as shown in Eqn. (2b), with $z_2$ being the propagation distance from sample to detector. The zeroth ($M_0$) and first moments ($M_1$) refer to the total beam intensity and centre of mass positions, respectively. While this is the data processing method used throughout the present work, we note that other methods (deconvolution, Gaussian perturbation etc.) provide comparable results[26]. To comprehensively determine whether EI/BT DF imaging can directly measure X-ray scattering signals as described by classical scattering theory, and how the measured signals vary according to beam, imaging system and sample parameters, we present data obtained at the European Synchrotron Radiation Facility (Grenoble, France) in which a single beamlet of variable vertical size and energy probes ensembles of monodisperse spheres. The spheres are of various radii (straddling the different beamlet dimensions), arranged in layers of varying thicknesses. Sphere ensembles were encased in gel wax slabs, and attenuation signals were fixed by maintaining a constant gel wax thickness through the inclusion of sphere-free slabs: regardless of the number of sphere-containing slabs, X-rays always traversed a total of five slabs. The microspheres were made of soda-lime glass ($\rho$ = 2.49 ± 0.04 gcm$^{-3}$), and the mass of each layer encased in gel wax was 0.0417 ± 0.0006 g, meaning that the estimated number of spheres in each layer decreases from 5.5×10$^7$ for the smallest (8 μm) diameter to 2.6×10$^3$ for the largest (233 μm). While the number of spheres changes with radius, the thickness of soda-lime glass traversed by x-rays is expected to be relatively constant across sphere sizes. The beamlet size is changed by varying the aperture in the pre-sample slit (see methods). A subset of measurements was repeated using a conventional x-ray source, to demonstrate the generality of the obtained results and their applicability outside specialised synchrotron facilities.

There are four parameters which may affect the DF signal in our (multiple-scattering) experiment:

1) the beamlet width;
2) the characteristic feature size of the sample;
3) the sample thickness;
4) the X-ray energy.

Clearly the sample material would also affect the DF signal but in our experiment is kept constant. The interplay between the first three parameters (at 60 keV) is explored in Fig 2, while energy is discussed in the extended data.

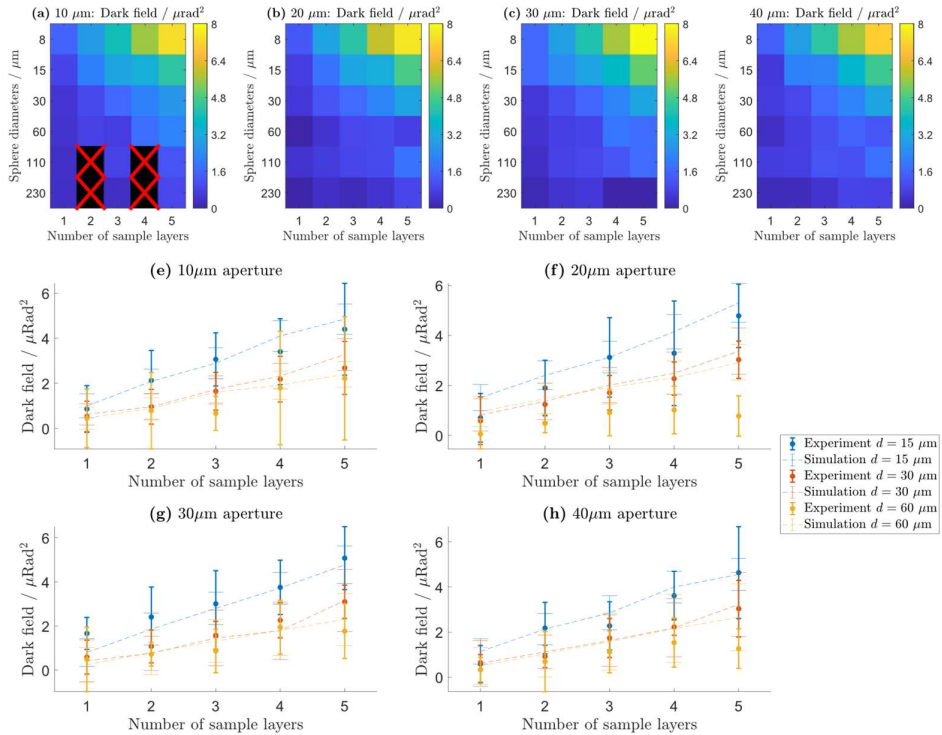

*Figure 2 DF signal vs sphere size, sample thickness and beamlet size. Panels (a-d) show experimental DF values for beamlet sizes of 10, 20, 30 and 40 µm (respectively); blocks marked with crosses indicate missing data. Sphere size and number of layers (used to create different sample thicknesses) are reported on the horizontal and vertical axes, respectively. Panels (e-h) show comparisons between (Monte-Carlo) model (dashed lines) and experiment (dots) for the same range of beamlet sizes. All plots show selected results vs. sample thickness for sphere diameters of 15 µm (blue), 30 µm (red) and 60 µm (yellow). Error bars represent one standard deviation over the measurements, and therefore include the variation in sphere distribution within the considered region-of-interest.*

Fig 2(a-d) make it evident that the DF signal increases with sample thickness (the number of traversed layers containing microspheres)[27,28] and with decreasing sphere size, in part because, as explained above, effort was made to keep constant the fill factor by volume which led to a total number of spheres inversely proportional to the cube of their diameter. While a similar dependence on sample thickness and sphere size has been observed before[29], a first novelty observed here is the independence from the size of the beamlets, despite the fact that resolution in EI/BT is defined by the aperture width[30]. As well as from the heatmaps in Fig 2(a-d), this can be clearly observed in the graphs of Fig 2(e-h). This is an important result because, together with EI/BT's independence from the spacing between adjacent beamlets (shown in previous work[22,30] and further discussed below), it demonstrates the independence of the retrieved scattering signal from system-specific parameters such as the autocorrelation length (defined as $\xi = \lambda d/p$, where $\lambda$ is the x-ray wavelength, $d$ the sample-to-detector distance, and $p$ the period of the mask or grating employed).

Another important result evident from Fig 2(e-h) is the agreement between model and experiment across the widely explored experimental conditions. Taken together with the agreement between model and scattering theory shown above, this demonstrates the ability of EI/BT to perform direct scattering measurements. For increased generality, a subset of samples were also scanned with a conventional lab system (Extended Data Fig 1) i.e., with an extended, polychromatic source. The quantitative agreement observed between model and experiment persists and demonstrates that the same scattering measurements can be performed with conventional lab sources. However, in this case the polychromaticity of the measurements requires the adoption of appropriate spectral weighting approaches when modelling[17].

Measurements were also taken at 90 and 120 keV to demonstrate EI/BT's robustness against increasing x-ray energy, which opens the way to applications to higher $Z$ or denser materials while remaining quantitative, as proven by the good agreement between model and experiment observed in all cases (Extended Data Fig 2). This also provided an opportunity to study the behaviour of the DF signal vs. energy, with the proven reliability of the simulation allowing us to fill in more energies than the three (60; 90; 120 keV) that were experimentally tested. Results are shown in Extended Data Fig 3. The best fit indicates proportionality to $E^{-3.7}$, although the previously reported[27,22] $E^{-4}$ dependence also leads to a reasonably good fit with a reduction in the coefficient of determination $R^2$ from 0.998 to 0.988, meaning an $E^{-4}$ dependence cannot be excluded. It should also be noted that $E^{-3.89}$ dependence was previously reported[31], hence this could be an area that requires further investigation. The independence of the EI/BT DF signal from the beamlet size becomes apparent when one considers the way in which it is retrieved in Eqn. (2(b)). By subtracting the beamlet width without the sample, its influence is eliminated. Conversely, in grating-based methods, the DF signal is measured as a reduction in fringe visibility[29,32–34] $V^S/V^R$, extracted either via fringe-scanning methods[35] or Fourier analysis[36], with $V = (I_{max} - I_{min})/(I_{max} + I_{min})$ with $I_{max}$, $I_{min}$ the maximum and minimum intensities observed in a sinusoidal pattern and indices S and R having the same meaning as above. This makes the measured signal dependent on the system's autocorrelation length, $\xi$, regardless of whether absorption rather than phase gratings are used[37]. This significant difference between the two methods is exemplified in figure 3.

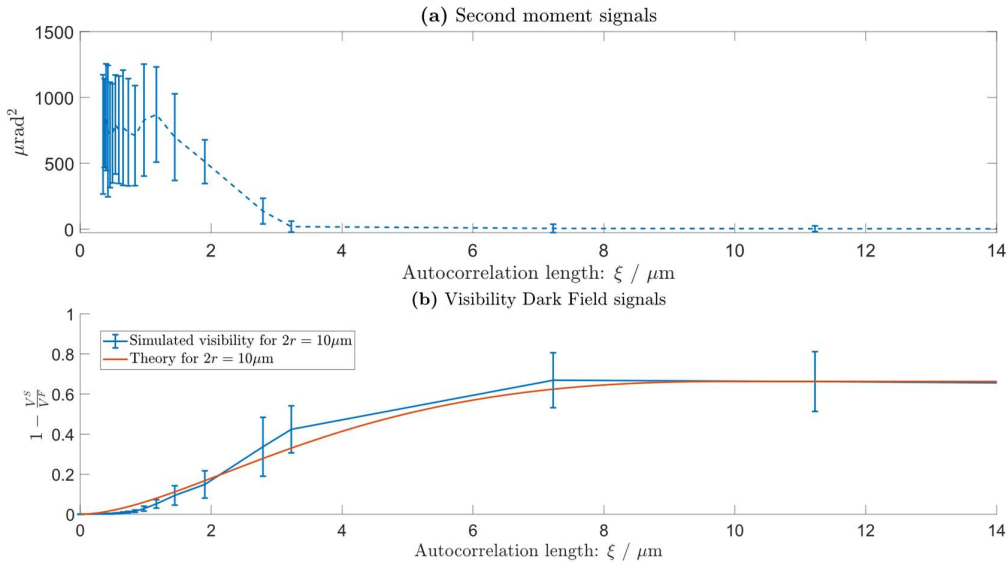

***Fig. 3 DF signal in EI/BT (a) and grating-based methods (b).*** *An ensemble of 10 µm diameter spheres was simulated at 20 keV and the resulting DF signals are retrieved as differences between 2$^{nd}$ moments and visibility reduction in (a) and (b), respectively. In both cases, the DF signal is plotted against the autocorrelation length $\xi$, which is initially varied by decreasing the grating period from 80 µm to 20 µm while keeping the aperture size (10 µm) and sample to detector distance (0.5 m) constant. To simulate periods smaller than 20 µm, half-period*

*apertures were used (i.e., Ronchi gratings were used for $\xi \geq 1.4$ µm), with the resulting beamlets beginning to overlap significantly. When the period reaches 4 µm ($\xi = 3.2$ µm), corresponding to an effective transition to grating interferometry, the autocorrelation distance is increased by increasing the grating to detector distance in half-Talbot distance steps. Error bars represent one standard deviation over the simulated region-of-interest. The theoretical behaviour as predicted by Strobl is overlaid as a red curve to the simulated points in (b), showing good agreement.*

The key aspect to note in Fig 3 is that, as predicted by Strobl[33] and Lynch[29], in grating-based methods the DF signal "kicks in" and remains stable when $\xi$ becomes larger than the average size of the scatterers, as can be seen by the curve "plateauing" around $\xi = 10$ µm in Fig. 3(b). It should be noted that, while most GI literature plots the DF signal as $V^S/V^F$, it is plotted here as 1-Vr/Vs to highlight the fact that the DF signal goes to zero for small values of $\xi$, and stabilises to its final (asymptotic) value only when $\xi$ exceeds the size of the scatterers. Conversely, with EI/BT, the scattering value is effectively retrieved at extremely small values of $\xi$, and is unaffected by it. What happens around $\xi = 1.4$ µm is that the specific geometry of the system causes excessive beamlet overlap, which prevents a "moment-type" retrieval as it becomes impossible to integrate the beamlets over their full extent. Other methods (e.g. Gaussian perturbation[34]) can be used to perform an effective DF retrieval at (moderately) larger $\xi$ values; however, the whole point of EI/BT is to keep the beamlet sufficiently separated from each other so as to allow a retrieval approach that focuses on the individual beamlet (considered as a separate, independent entity), rather than on the analysis of the pattern as a whole. Indeed, as beamlet overlap increases, the modulation ultimately becomes sinusoidal, and EI/BT becomes indistinguishable from grating-based approaches[38]. With this in mind, it is important not to misinterpret the graph in Fig. 3(a): the decrease of the retrieved DF signal for $\xi > 1.4$ does not indicate a dependence of the retrieved DF on $\xi$, but rather that EI/BT systems with such high $\xi$ values should not be built, as they would effectively not be EI/BT systems anymore since they do not allow the individual analysis of each beamlet. In other words, in the described settings an EI/BT system must have a $\xi < 1.4$ and, in this $\xi$ range, scatter retrieval is not affected by the specific value of $\xi$. This also means that the concept of $\xi$ does not really apply to EI/BT, and indeed e.g. increasing *d* and decreasing *p* do not have the same effect on an EI/BT system.

Models for grating-based DF were also used to determine a sample's "scattering power" as a function of the scatterer's diameter, for example through a "DF extinction coefficient" (see e.g. Fig 4 in Lynch 2011[29]). The models described in this paper allow doing the same for EI/BT, revealing that in this case a monotonic function is obtained (see Extended Data Fig 4), because of the discussed independence from specific imaging system parameters. For completeness, the ability of the same (wave-optics) model to reproduce the results obtained using a standard phase gratings setup is also shown (Extended Data Fig 5).

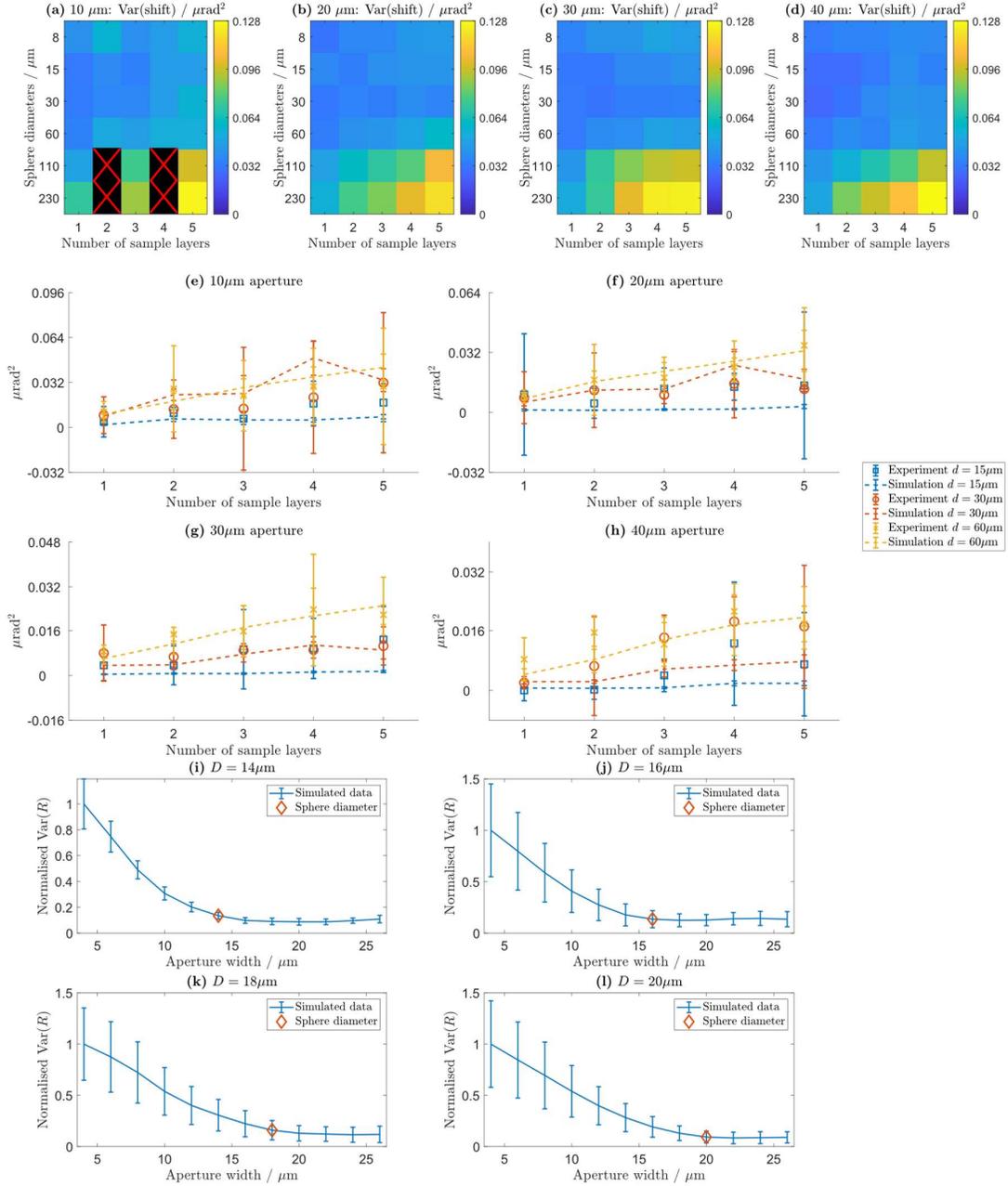

*Fig 4 Variance of refraction (VR)*. Panels (a-d) show the same heatmaps as Fig. 2(a-d) for the VR instead of the DF signal, for beamlet sizes of 10, 20, 30 and 40 $\mu m$ (respectively); blocks marked with crosses indicate missing data. Sphere size and number of layers (used to create different sample thicknesses) are reported on the vertical and horizontal axes, respectively. Panels (e-f) show comparisons between model (dashed lines) and experiment (dots) for the same range of beamlet sizes. All plots show selected results vs. sample thickness for sphere diameters of 15 $\mu m$ (blue), 30 $\mu m$ (red) and 60 $\mu m$ (yellow). Finally, panels (i-l) show simulated data vs. beamlet size for scatterers of various diameter, showing how the plots plateau when the size of the scatterer matches that of the beamlet used to probe the sample, in analogy with the behaviour of DF as a function of $\xi$ in grating-based methods. Error bars represent one standard deviation over either the experimentally considered or simulated region-of-interest, and as such they include the variation in sphere distribution therein.

It has been argued that the dependence of "visibility-based" DF retrieval on $\xi$ can be used to determine the size of the scatterers, although this requires sweeping $\xi$ which can be experimentally challenging (the most practical method probably being based on changing the distance between two phase gratings[39]). This is not possible in EI/BT experiments in which, so long as the beamlets are kept physically separated, a constant value directly linked to the sample's scattering power, which is unaffected by system parameters, is returned. Although models exist that allow retrieving the scatterers' size so long as more information is extracted from the scattering function on top of its variance[25], the fact that the refraction signal is sensitive to objects larger that the system's spatial resolution can be exploited for this purpose.

In analogy with Fig 2(a-d), Fig 4(a-d) shows the (experimental) behaviour of the variance of the refraction signal (VR) as a function of the scatter size and thickness for various beamlet sizes. The signal is obtained by applying a standard deviation filter with a nine-by-nine pixel window, on a flat-field corrected refraction image, squaring it, and averaging the same ROI as for the DF measurements. As can be seen, the behaviour is comparable to that of DF vs. sample thickness (i.e. number of scattering layers), but opposite vs. scatterer size, with the maximum values observed for the larger spheres. This can be understood intuitively as features below the spatial resolution of the system do not produce distinct refraction signals. Also, the VR signal is not independent from the beamlet size, as made clear also by the graphs of Fig 4(e-h), which also show that the VR signal can be modelled with the same tools used for the DF signal, obtaining a comparable agreement with the experimental results. This allows us to use the model to explore the VR signal of monodisperse spheres as a function of a varying beamlet size, with a denser sampling than would be practically feasible in a real experiment. The results (Fig. 4(i-l)) show a clear plateauing of the signal when the size of the beamlet reaches that of the scatterers, similarly to (visibility-based) DF vs. $\xi$ in grating-based methods. As a consequence, it could be used in the same way – e.g. by using a pre-sample mask with small apertures and translating the mask in aperture increments, the separate images then being summed will create virtual beamlets of larger widths.

We are effectively proposing the VR as a new contrast channel which provides an "equivalent" to DF for randomly distributed features larger than the spatial resolution of the system, with the same characteristics of e.g. increasing linearly with sample thickness for a given distribution of scatterers. This had already been used heuristically to determine the degree of porosity in composite materials[40], and is formalised here as an additional, quantitative contrast channel. The agreement with the theoretical models allows using the latter to explore the behaviour of VR vs. e.g. sample thickness and x-ray energy, with exemplar results reported in Extended Data Fig 6. In particular, the linear behaviour of VR vs sample thickness demonstrates its suitability to tomographic implementations.

The simultaneous availability of DF and VR signals through a single scan of the same sample thus offers a method for the quantitative characterisation of the sample's microstructure, above and below the system's spatial resolution, with direct access to the sample's scattering function unaffected by specific parameters of the imaging system.

# Methods

## Synchrotron experiments

Synchrotron experiments were carried out at beamline ID17 of the European Synchrotron Radiation Facility (ESRF). The radiation source is a wiggler placed at approximately 145 m from the experimental hutch. The beam is monochromatized by a bent Laue crystal and energies of 60, 90 and 120 keV were selected. An EI/BT setup was installed in the experimental hutch, a side view schematic of which is shown in Fig. 5. The beam (in yellow) is vertically thin but extends to several cm into the plane of the drawing so as to cover the entire imaged samples.

A tuneable Huber slit ($A_1$) was used as a pre-sample mask, so that samples could be inspected with variable (vertical) beam sizes – sizes of 10, 20, 30 and 40 µm were used. A much larger (300 µm) slit ($A_2$) was placed in front of a 22.4 µm pixel detector (PCO Edge 5.5 sCMOS detector), mounted on a vertical translation stage. This allows rapid switching between the EI and the BT configuration. By letting the beam entirely through $A_2$ (as in the schematic of Fig. 5), the beamlet is directly resolved by the detector pixels (BT mode). By moving $A_2$ downwards or upwards so that the beam is partially intercepted by one of its edges, the EI configuration is obtained. In this latter case, multiple sample scans, with slightly different degrees of beam or edge overlap, are required to enable the retrieval of the beam's vertical width; in principle, a much larger (e.g. >300 µm) detector pixel could be used, but in this case the 22.4 µm pixels were binned together for simplicity. In both cases, the sample is scanned vertically through the beam to cover the desired region-of-interest.

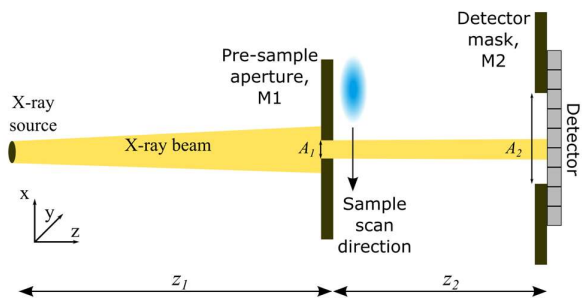

***Fig. 5 EI/BT setup***. *A slit is placed immediately before the sample; its aperture size $A_1$ is tuneable thus allowing varying the size of the beamlet hitting the sample. A second slit with a larger aperture $A_2$ is mounted in front of a high resolution detector, with images being obtained by scanning the sample through the beam. Both BT and EI configurations are possible with this setup: the former by analysing the beamlet directly with the detector pixels as shown, the latter by summation of vertical pixels and translation of M1 in the $x$ direction.*

## Laboratory experiments

These were performed with a standard laboratory EI system, previously described in Astolfo 2022[41]. Briefly, a COMET MXR-160/11 tungsten anode x-ray source had its focal spot limited to 70 μm in the horizontal direction by a slit, with the vertical dimension left at the original ~400 μm. For the experiments presented in this paper, it was operated at 60 kVp and 7.7 mA. Two gratings (fabricated by Microworks, Karlsruhe, Germany) with 21 and 28 μm parallel, vertical apertures arranged in 75 and 98 μm periods were used as pre-sample and detector masks, respectively. Both had approximately 300 μm thick gold septa on a 1 mm thick graphite substrate, and they were placed 1.60 and 2.06 m downstream of the source, respectively. The samples were scanned horizontally (i.e. orthogonally to the gratings' slits) immediately downstream of the pre-sample mask with a Newport (Irvine, CA) translator at a speed of 39.1 μm s$^{-1}$, and a 100 μm pixel photon-counting detector (XC-FLITE FX2, Direct Conversion, Danderyd, Sweden – now part of Varex), acquiring at 1 frame per second, was placed 4 cm downstream of the detector mask.

## Wave optics and Monte Carlo X-ray propagation models

The rigorous multi-slice wave optical simulation code developed by Munro[42] was adapted to simulate the EI/BT synchrotron experimental setup, as well as the simulations highlighting differences and similarities between EI/BT and grating interferometry. Planar wavefronts are propagated via the Fresnel propagator, with sample effects being applied through multiplication of the wavefront by the samples' complex transmission functions. Due to the large distances between source, sample, and detector, each 'layer' of spheres could be simulated as a single projection. To keep computation times manageable, only small sample areas (0.6 × 2.1 mm$^2$) were simulated; this is a particular advantage when considering that the field of view available in such simulations becomes more costly with increasing photon energy in order to preserve robustness against aliasing. Finally, upon reaching the detector, beam divergence is accounted for by application of the Fresnel scaling theorem.

The McXtrace neutron / X-ray scattering engine developed by Knudsen et al[43] was used to simulate the synchrotron and lab-based experimental setups. Refraction is implemented in three dimensions via Snell's law, whereby photons refract or reflect as they encounter sample boundaries[24]. Absorption is calculated by adjustment of photon 'weights' according to the Beer-Lambert law, and in this way all photon paths that are computed contribute to the final image. In the polychromatic lab setting, photons are assigned energies according to the CdTe efficiency-weighted spectrum produced by TASMICS[44].

Virtual samples of randomly arranged and tightly packed spheres were created for use in both the simulation frameworks described above via the Advancing Front algorithm described by Valera et al[45]. The average volumetric fill of simulated spheres was 42 ± 8%.

## Sample preparation

Melted gel wax was poured into a 2 cm × 2 cm × 1 cm mould, with a 1 cm × 1 cm × 0.5 cm cuboid void extending inwards from one of the wide faces, and allowed to cool. Mass-controlled amounts of calibrated microspheres were added to the voids and manually vibrated until their distribution was approximately uniform. Melted gel wax was then poured into the void to seal the microspheres in place, with the excess gel wax being scraped away with a glass microscope slide moving parallel to the existing gel wax surface to minimise any variation in gel wax (i.e., background material) thickness. As the prepared samples cooled, they were monitored for any signs of air having been trapped in the gel wax, as such pockets of air would produce their own phase signals- any such exhibiting samples were discarded to ensure that only the calibrated microspheres contributed to the final image contrasts.

## Acknowledgements

This work was supported by the EPSRC (Grant EP/T005408/1). AO and CKH are supported by the Royal Academy of Engineering under the "Chairs in Emerging Technologies" and "Research Fellowships" schemes, respectively. PRTM is supported by the Royal Society under the "University Research Fellowships" scheme. The authors thanks Dr Heinz Amenitsch (Elettra Synchrotron Light Source ,Trieste, Italy) for useful discussions.

# Extended Data

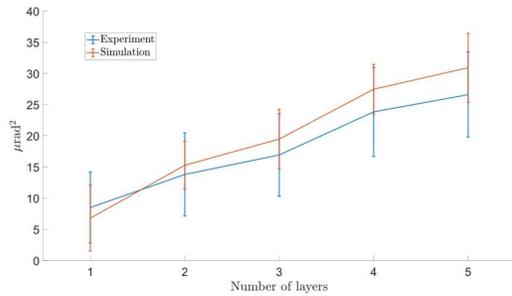

**Extended Data Fig 1:** simulated vs experimental DF signal acquired with a laboratory x-ray system, using a conventional reflection target source operated at 60 kVp (see methods section for details), for an increasing number of layers containing 15 μm diameter microspheres (corresponding to the blue lines and dots in the graphs in Fig 2(e-h) of the main article). As can be seen, a comparable level of agreement as in Fig. 2 is obtained. To reproduce the polychromatic results, simulated values for individual energies are weighted according to the spectral entries of the used (W) x-ray spectrum. Error bars represent one standard deviation over the measurements, and therefore include the variation in sphere distribution within the considered region-of-interest. This also explain why they are larger in the experimental case, as it is easier to generate a reasonably uniform sphere distribution in a synthetic model compared to a real sample.

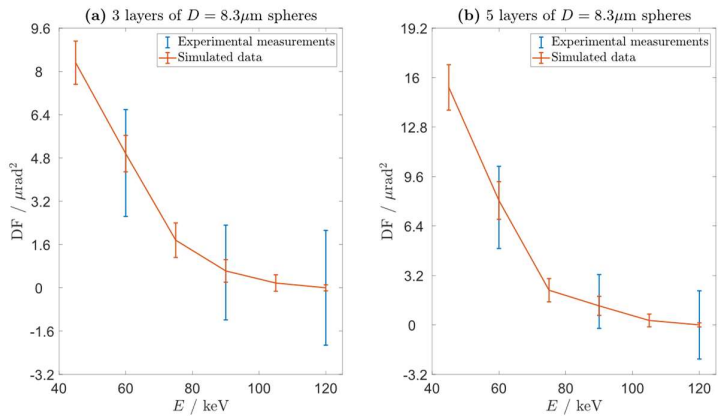

**Extended Data Fig. 2:** comparison between model and experiment vs. varying x-ray energy for 3 (a) and 5 (b) layers of the 8 μm diameter spheres. Experimental data were collected at 60, 90 and 120 keV; the model was used to create corresponding DF signals at the same energies as well as at 45, 75 and 105 keV to better highlight the trend of DF signal vs. x-ray energy. As can be seen, a good agreement is obtained, which highlights the reliability of the model also at high x-ray energies. Error bars correspond to the standard deviation observed in the analysed regions-of-interest.

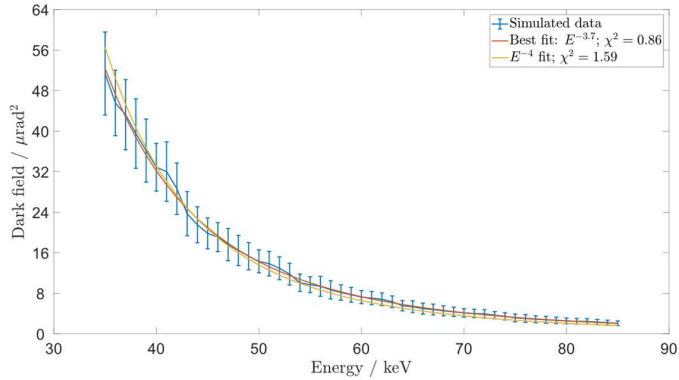

**Extended Data Fig. 3:** fitting the data shown in Extended Data Fig. 2 yields an energy dependence of $E^{-3.7}$, which is in slight disagreement with the trend previously reported in the literature ($E^{-4}$: the square of the energy dependence of δ). To refine this analysis, the model was used to create more finely sampled data in the lower energy window explored in previous literature (every 1 keV, blue markers with error bars, corresponding to the standard deviation observed in the simulated regions-of-interest) and the fitting procedure was repeated, still yielding $E^{-3.65}$ as the best fit (solid red line). However, imposing an $E^{-4}$ dependence (solid yellow line) still yields a reasonable fit, leading to only a small reduction in the determination coefficient (from 0.998 to 0.988). This study therefore cannot rule out a $E^{-4}$ dependence with statistical significance; however, it should be noted that exponents smaller than 4 were previously reported in the literature (see main text).

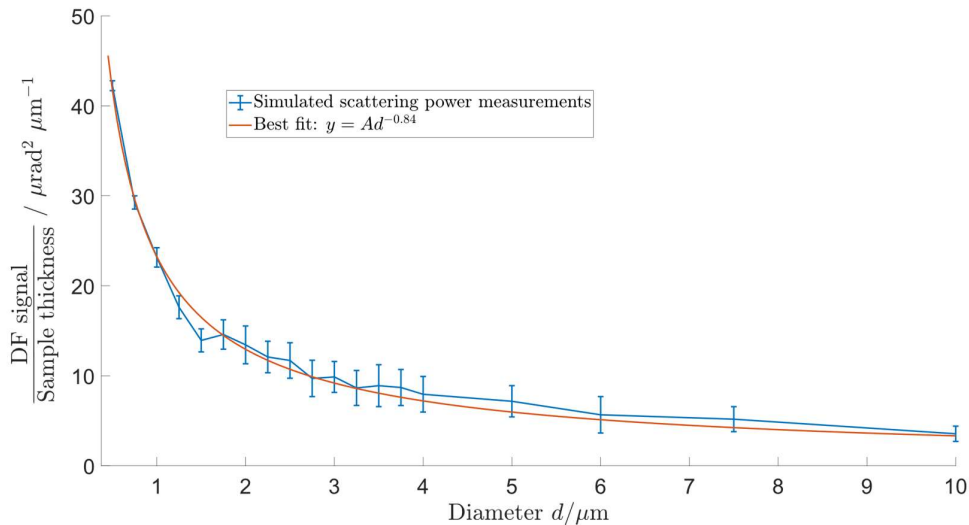

**Extended Data Fig. 4:** "scattering power" in EI/BT. As done in previous studies on grating-based methods, this assumes the DF signal is proportional to $tf(d)$, where $t$ is the total thickness of the traversed material and $f(d)$ a parameterised function of the scatterers' diameter, $d$. The above graph shows this DF signal power, extracted from simulated ensembles of spheres with constant packing density and varying diameters, illuminated by a 10 μm beamlet at 20 keV. The blue points are the extracted data, and their error bars represent one standard deviation over the simulated region-of-interest; the red line describes a well-fit monotonically decreasing curve, namely a $d^{-0.84}$ power function.

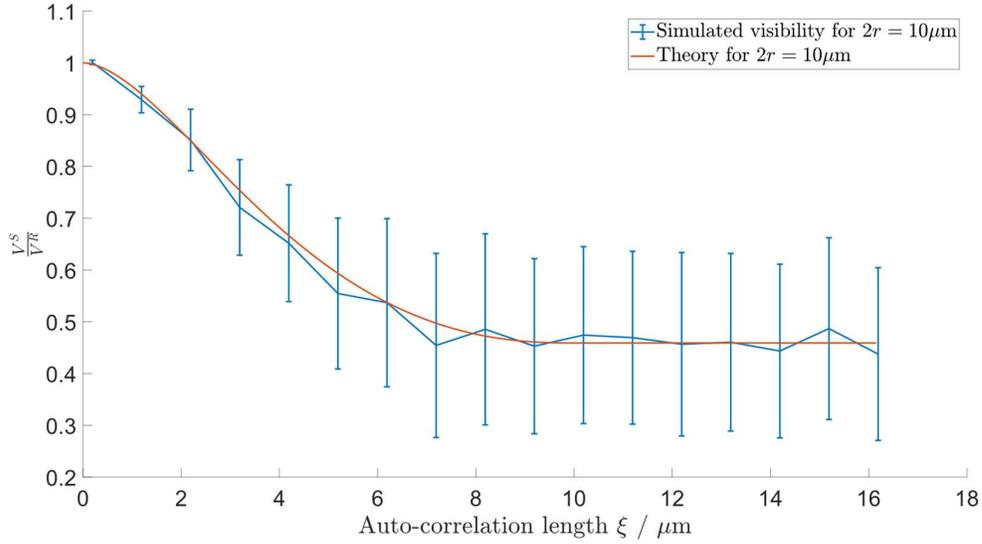

**Extended Data Fig. 5:** behaviour of the visibility-based DF signal in grating interferometry reproduced using the (wave-optics) model described in the main article. A π phase grating with 1 μm period is placed 147.5 m downstream of a 20 keV point source, an ensemble of $d = 10$ μm spheres is placed 1 cm downstream of the grating, and visibility measurements are collected at increasing Talbot orders, corresponding to increased $\xi$ values. Error bars correspond to the standard deviation observed in the simulated regions-of-interest. The red line represents the theoretical behaviour derived by Strobl, showing good agreement with the simulated data.

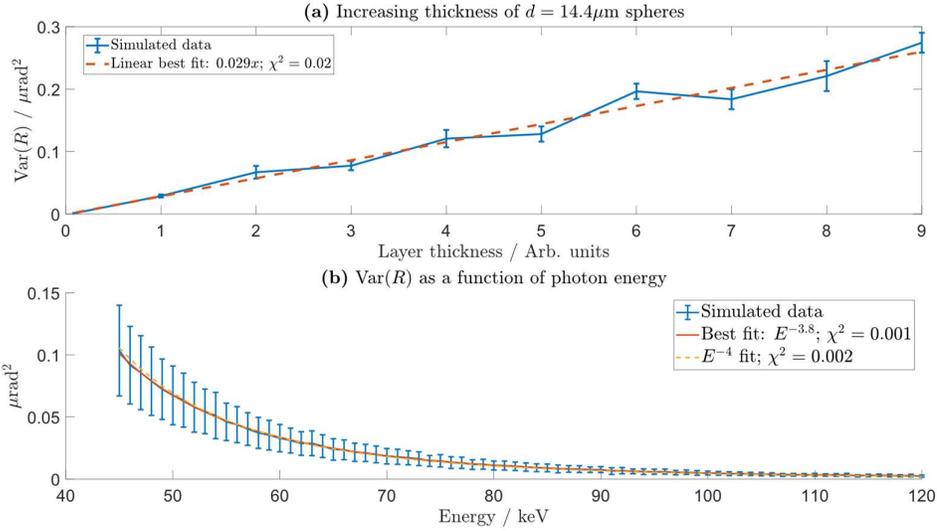

**Extended Data Fig. 6:** dependence of the variance of refraction (VR) signal on sample thickness (a) and x-ray energy (b), obtained from simulated data following the experimental validation of the model shown in Fig. 4(e-h) of the main manuscript. The signal is linear with sample thickness (a), and decreases approximately with $E^{-3.8}$ (b), with the data being compatible with an $E^{-4}$ dependence (with a marginal reduction in the coefficient of determination from 0.999 to 0.997). In this sense, the behaviour is the same as for the DF signal. Data were generated for 60 μm diameter spheres illuminated by a 20 μm beamlet at 60 keV; error bars represent the standard deviation of the signal in the simulated regions-of-interest.